\documentclass[conference]{IEEEtran}
\IEEEoverridecommandlockouts
\usepackage{cite}
\usepackage{amsmath,amssymb,amsfonts}
\usepackage{algorithmic}
\usepackage{graphicx}
\usepackage{textcomp}
\usepackage{xcolor}
\def\BibTeX{{\rm B\kern-.05em{\sc i\kern-.025em b}\kern-.08em
    T\kern-.1667em\lower.7ex\hbox{E}\kern-.125emX}}
\begin{document}

\title{Deep Learning based NAS Score and Fibrosis Stage Prediction from CT and Pathology Data}

\author{\IEEEauthorblockN{Ananya Jana*}\thanks{* equal contribution}
\IEEEauthorblockA{\textit{Computer Science Dept.} \\
\textit{Rutgers University}\\
New Brunswick, USA \\
ananya.jana@rutgers.edu}
\and
\IEEEauthorblockN{Hui Qu*}
\IEEEauthorblockA{\textit{Computer Science Dept.} \\
\textit{Rutgers University}\\
New Brunswick, USA  \\
hui.qu@cs.rutgers.edu}
\and
\IEEEauthorblockN{Puru Rattan}
\IEEEauthorblockA{\textit{Division of Gastroenterology and Hepatology} \\
\textit{Rutgers Robert Wood Johnson Medical School}\\
New Brunswick, USA \\
puru.rattan@rutgers.edu}
\and
\IEEEauthorblockN{Carlos D. Minacapelli}
\IEEEauthorblockA{\textit{Division of Gastroenterology and Hepatology} \\
\textit{Rutgers Robert Wood Johnson Medical School}\\
New Brunswick, USA \\
minacacd@rwjms.rutgers.edu}
\and
\IEEEauthorblockN{Vinod Rustgi}
\IEEEauthorblockA{\textit{Division of Gastroenterology and Hepatology} \\
\textit{Rutgers Robert Wood Johnson Medical School}\\
New Brunswick, USA \\
vinod.rustgi@rutgers.edu}
\and
\IEEEauthorblockN{Dimitris Metaxas}
\IEEEauthorblockA{\textit{Computer Science Dept.} \\
\textit{Rutgers University}\\
New Brunswick, USA  \\
dnm@cs.rutgers.edu}
}

\maketitle

\begin{abstract}
Non-Alcoholic Fatty Liver Disease (NAFLD) is becoming increasingly prevalent in the world population. Without diagnosis at the right time,  NAFLD can lead to non-alcoholic steatohepatitis (NASH) and subsequent liver damage. The diagnosis and treatment of NAFLD depend on the NAFLD activity score (NAS) and the liver fibrosis stage, which are usually evaluated from liver biopsies by pathologists. In this work, we propose a novel method to automatically predict NAS score and fibrosis stage from CT data that is non-invasive and inexpensive to obtain compared with liver biopsy. We also present a method to combine the information from CT and H\&E stained pathology data to improve the performance of NAS score and fibrosis stage prediction, when both types of data are available. This is of great value to assist the pathologists in computer-aided diagnosis process. Experiments on a 30-patient dataset illustrate the effectiveness of our method. 
\end{abstract}


\section{Introduction}
Nonalcoholic fatty liver disease (NAFLD) is now the most common form of chronic liver disease in the world. The prevalence of this disease has been estimated to be over 25\% in the general worldwide population~\cite{asrani2019burden}. 
This disease encompasses a spectrum of changes in the liver related to fat deposition. The changes range from non-alcoholic fatty liver (NAFL) with simple steatosis ($>5\%$ liver fat content) with minimal or no inflammation, to a progressive form of the disease called non-alcoholic steatohepatitis (NASH). NASH is characterized by steatosis, inflammation and hepatocellular injury, with eventual progression to various stages of fibrosis~\cite{bedossa2017pathology}. As the progressive form of the disease, NASH is associated with increased morbidity and mortality; therefore, determination of this disease is important during diagnosis. Liver biopsy is the gold standard for determining the fibrosis stage and diagnosing NASH from NAFLD activity score (NAS) to differentiate it from simple steatosis. However, the cost of an invasive procedure such as liver biopsy combined with the possibility of complications such as bleeding, infections, and rarely death, rule out its routine use in clinical practice~\cite{chalasani2018diagnosis}. In addition, intra-observer variance along with sampling variance add significant error to the manual interpretation of liver biopsy histopathology~\cite{ratziu2005sampling,pournik2014inter}. Therefore, it is of great value to develop computational methods for NAS score prediction and fibrosis staging from non-invasive imaging modality data. Besides, when the biopsy data is available, it is meaningful to build computational models to predict scores from the biopsy data, which can assist pathologists in the diagnosis process by saving time and reducing the unreliability of less-experienced doctors.

\begin{figure*}[t]
\centering
\includegraphics[width=0.8\textwidth]{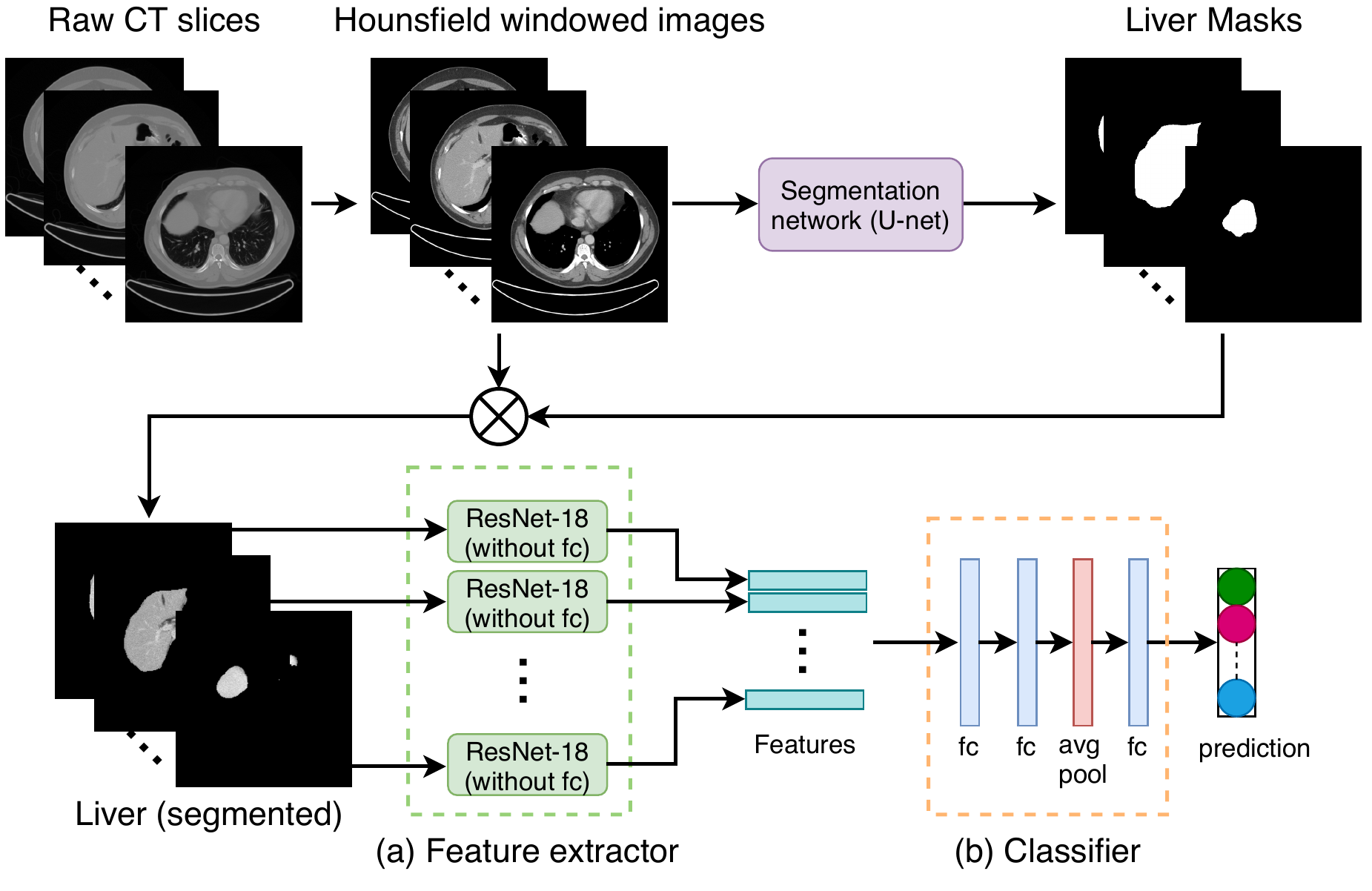}
\caption{CT data preprocessing and Baseline network structure.}
\label{fig:overview}
\end{figure*}

Recently, deep learning has been immensely successful in image analytics tasks such as image classification~\cite{russakovsky2015imagenet} and semantic segmentation~\cite{cordts2016cityscapes}. The ability to learn meaningful features automatically makes deep learning superior to traditional machine learning approaches. 
There have been many deep learning based works on fibrosis staging and NAS score prediction. These methods cover various types of data, such as CT scans~\cite{yasaka2018deep,choi2018development,obmann2018ct}, Magnetic Resonance Imaging (MRI)~\cite{yasaka2017liver,venkatesh2015non,rustogi2012accuracy}, pathology images from liver biopsy~\cite{fu2018segmentation,yu2018deep,heinemann2019deep} and others~\cite{treacher2019deep,gatos2019temporal}. CT and MRI scans are non-invasive,  thus it is beneficial for patients if one can obtain accurate fibrosis stage and NAS scores from these scans. Compared with CT and MRI data, pathology images are more informative about the disease, and are used as gold standard for diagnosis. For fibrosis staging, Yasaka et al.~\cite{yasaka2018deep} explored the use of a deep convolutional neural network (CNN) for staging liver fibrosis on magnified CT images of the liver surface. They concatenated the age and sex information of the subject at one of the fully connected layers. The scores found from the model were moderately correlated with the liver fibrosis stage as found from histopathology. Jin et al.~\cite{choi2018development} proposed a fibrosis prediction mechanism where the liver region in the image was first segmented using a segmentation network and then classified using a CNN. Yu et al.~\cite{yu2018deep} investigated and compared the performance of liver fibrosis stage classification using different deep learning algorithms and other machine learning algorithms. Fu et al.~\cite{fu2018segmentation} explored fibrosis identification with the help of image segmentation. But their method didn't predict the exact fibrosis stage. Heinemann et al.~\cite{heinemann2019deep} trained a CNN to predict fibrosis by using histology images at different scales. For NAS score prediction, pathology data is more often used. Heinemann et al.~\cite{heinemann2019deep} trained a CNN to predict the three individual NAS scores - NAS steatosis, NAS lobular inflammation, NAS ballooning.  NAS score prediction from CT images  has rarely been explored. Hence, an important question is: can we predict NAS scores directly from CT images? Besides, when the paired pathology data is also available, can we utilize both CT and pathology data to improve the prediction performance? In practice, the diagnoses using CT scans and pathology slides are performed by radiologists and pathologists, respectively. If we can combine the information from both types of doctors (i.e., train a model that collect information from both types of data), it could be beneficial to the final results.

\begin{figure*}[t]
\centering
\includegraphics[width=0.8\textwidth]{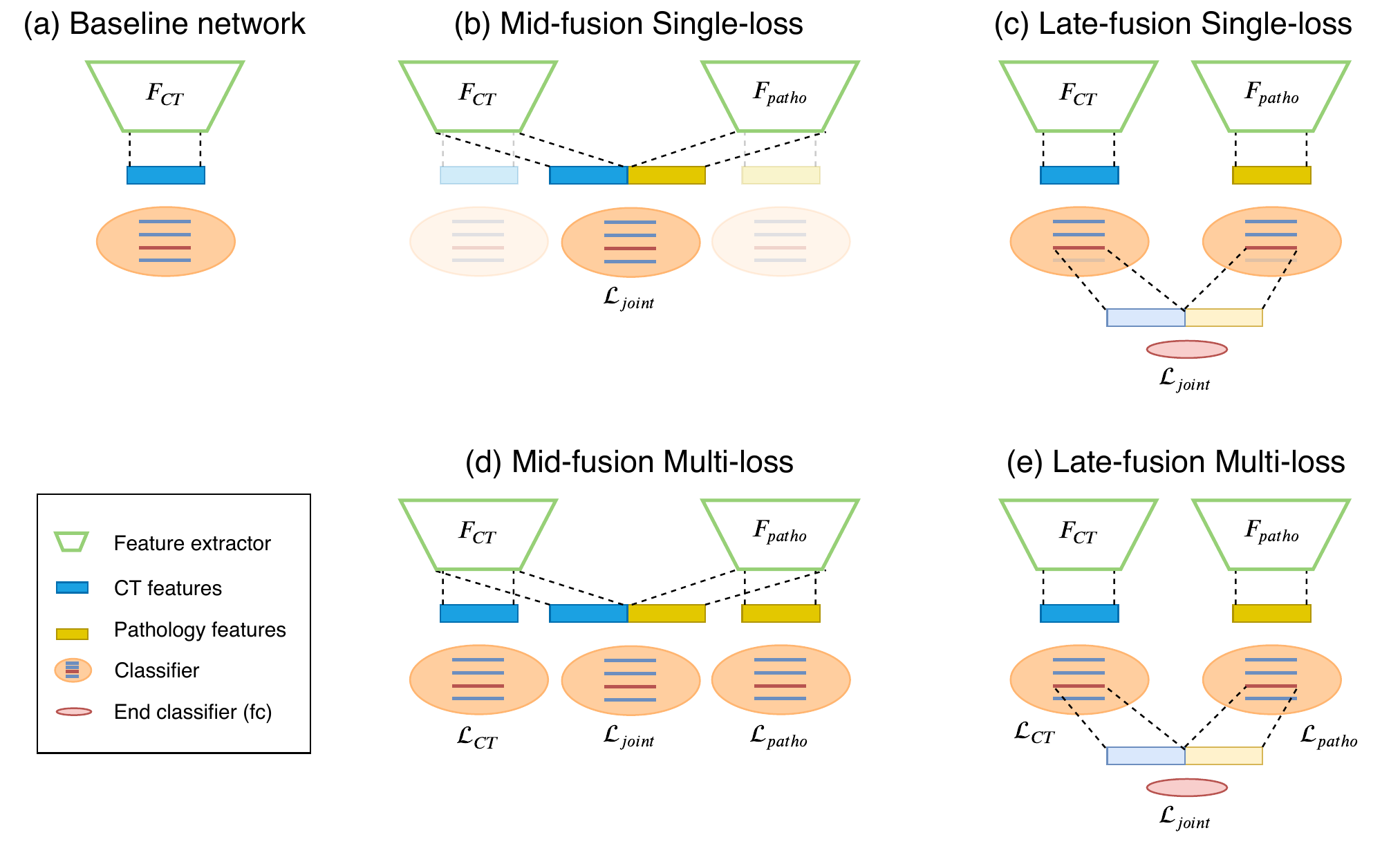}
\caption{Baseline network and four different joint networks.}
\label{fig:fusion}
\end{figure*}

In this work, we propose to predict the individual NAS scores (NAS steatosis, NAS lobular inflammation, NAS ballooning) directly from CT images using deep learning. The 3D CT volumes are divided into 2D slices, in which the liver portion is segmented. Features are extracted from these slices by a pretrained ResNet~\cite{he2016deep}, and then aggregated by a classifier for prediction. Besides, we also combine the information from CT scans and pathology images to further improve the performance, considering that they may contain different levels of information about the disease. The CT and pathology images are fed into two separate networks to produce features that are related to fibrosis or NAS. Then the two types of features are fused to get the final classification result. We explore different feature fusion strategies and loss functions. The proposed method can achieve a good performance on a 30-patient dataset that contains paired CT and pathology data. In summary, our main contributions include:
\begin{itemize}
    \item We propose a novel method to predict NAS scores directly from CT images.
    \item We design a network to predict fibrosis stage and NAS score from CT and pathology images, and we achieve better performance than any single type of data.
    \item As far as we know, we are the first to perform deep learning based NAS score prediction using CT images and  to use multiple-modality data for fibrosis staging and NAS score prediction.
\end{itemize}

\section{Method}
The overview of our proposed method is shown in Fig.~\ref{fig:overview} and Fig.~\ref{fig:fusion}. It consists of two main steps: (1) data preprocessing on CT scans and pathology slides, (2) network training using the preprocessed data. 

\subsection{Data preprocessing}
The raw CT and pathology data cannot be directly fed into CNNs for training. We perform preprocessing for each type of data.

\subsubsection{2D liver segmentation (CT data)} 
The raw CT scans are 3D volumetric data. To be compatible with the 2D pathology images during training, we extract 2D slices from each 3D scan. Hounsfield windowing with the range $[-200, 250]$ is performed on the 2D CT slices to increase the image contrast. As we only focus on the liver part in the CT images, we perform 2D liver segmentation to extract liver regions before the fibrosis stage and NAS score prediction tasks. All other pixels are set to zero to avoid the negative effect of other organs (shown in Fig.~\ref{fig:overview}). The segmentation model is first pretrained on the ISBI2012 dataset~\cite{cardona2010integrated}. Then we annotate a small part of the 2D slices in our dataset and fine-tune the segmentation model with the annotated images, utilizing transfer learning. We had made a train test split on the annotated liver dataset. The dice score of the segmentation network was 0.9421 on the test liver dataset. After liver segmentation, the slices with an average pixel value below a threshold (5 in our experiments) are discarded. This ensures that slices containing very small liver portions and those slices which do not contain liver portion at all are discarded. The number of CT slices after preprocessing is 2595.

\subsubsection{Patch generation (pathology data)}
The original pathology whole slide images (WSIs) have very high resolution (10,000 to $>$100,000 pixels on each dimension). It is impractical to put these images into a neural network directly. Besides, a large portion of pixels in each WSI belong to background and contain no information. 
It is better to remove such background areas during training, otherwise the useful information from tissue may be overwhelmed by uninformative pixels. Therefore, we extract non-overlapping image patches of resolution 224$\times$224 from each WSI at $5\times$ magnification. Patches that have a mean pixel value greater than 220 are considered as background and are removed. The number of pathology patches after preprocessing is 7775.

\begin{table*}[t]
    \centering
    \caption{Original and combined Fibrosis stage and NAS score distributions in the 30-subject dataset.}
    \begin{tabular}{|l|p{0.4cm}|p{0.4cm}|p{0.4cm}|p{0.4cm}|p{0.4cm}|p{0.4cm}||p{0.4cm}|p{0.4cm}|p{0.4cm}|p{0.4cm}||p{0.4cm}|p{0.4cm}|p{0.4cm}|p{0.4cm}||p{0.7cm}|p{0.7cm}|p{0.7cm}|}
    \hline
         & \multicolumn{6}{|c||}{Fibrosis} & \multicolumn{4}{|c||}{NAS steatosis} & \multicolumn{4}{|c||}{NAS lobular}  & \multicolumn{3}{|c|}{NAS ballooning}\\\hline
    Score & 0 & 1 & 2 & 3 & 3.5 & 4 &  0 & 1 & 2 & 3 & 0 & 1 & 2 &3 &0 &1 &2\\ \hline
    Original & 7 & 6 & 4 & 3 & 2 & 8 & 2& 9 & 11 & 8 & 9 & 10 & 8 & 3 & 8 & 11 & 11 \\ \hline
    Combined & 7 & \multicolumn{2}{c|}{10} & \multicolumn{3}{c||}{13} & 
          \multicolumn{2}{|c|}{11} & \multicolumn{2}{c||}{19} &  9 & 10 & \multicolumn{2}{|c||}{11} & 8 & 11 & 11 \\ \hline
    \end{tabular}
    \label{tab:score}
\end{table*}

\subsection{Baseline network for CT images}
This network is designed to predict the NAS scores based on CT data. The architecture, shown in Fig.~\ref{fig:overview} and Fig.~\ref{fig:fusion}(a), consists of a feature extractor and a classifier. The feature extractor aims to obtain a feature vector that represents the input 2D slice. We use the ResNet-18~\cite{he2016deep} (without the final fully-connected layers) as our feature extractor because of two reasons - (1) the more complex models would overfit our training data which is limited to 30 patients, (2) the training is done at patient level, i.e., all the pathology patches and CT 2D slices are fed to the network for training, making it hard to use larger models due to GPU memory limitations. The classifier consists of fully-connected layers and an average pooling layer. The first two fc layers have 512 and 128 neurons respectively, which further process the extracted features of all input CT slices of a patient. The subsequent average pooling layer is used to obtain the global feature vector from local feature vectors of 2D slices. The final fc layer predicts the classification result from the global feature. Each individual score (fibrosis, NAS steatosis, NAS lobular, NAS ballooning) is trained with one network.

The baseline network can be also used to predict NAS scores and fibrosis stage from histpathology images by just replacing the input 2D slices with the extracted patches from a WSI.

\subsection{Joint network for both data}
The structure of the joint network for both CT and pathology data are shown in Fig.~\ref{fig:fusion}. There are two separate baseline networks and a joint classifier. The two baseline networks take as input the CT images and the pathology images, respectively. The joint classifier takes as input the fused features from the two baseline networks, and outputs the prediction. We explore the effects of two different feature fusion strategies and two types of loss functions, resulting in four different architectures.

\subsubsection{Mid-fusion vs. late-fusion}
In the mid-fusion architectures (Fig.~\ref{fig:fusion}(b) and Fig.~\ref{fig:fusion}(d)), the outputs of the two baseline networks' feature extractors are concatenated and fed to the joint classifier, which has the same structure as the classifier in the baseline network. In this case, local features of all 2D CT slices and pathology patches are stacked together to produce a global feature representation of both types of data. 

In the late-fusion architectures (Fig.~\ref{fig:fusion}(c) and Fig.~\ref{fig:fusion}(e)), the fusion is done after the average pooling layers of the two baseline networks. That's to say, we concatenate the global features (1$\times$128 each) obtained from both types of data to form a single feature (1$\times$256). The fused global feature is sent to the joint classifier, which consists of a fully connected layer. 

\subsubsection{Single-loss vs. multi-loss}
For the single-loss architectures (Fig.~\ref{fig:fusion}(b) and Fig.~\ref{fig:fusion}(c)), we only compute the loss with regard to the output of the joint classifier, i.e., $\mathcal{L} = \mathcal{L}_{joint}$. It only cares about whether the prediction from both types of data is correct or not. 

For the multi-loss architectures (Fig.~\ref{fig:fusion}(d) and Fig.~\ref{fig:fusion}(e)), the final loss $\mathcal{L}$ is the summation of the joint loss and losses from each individual baseline network, i.e., $\mathcal{L} = \mathcal{L}_{joint}+\mathcal{L}_{CT}+\mathcal{L}_{patho}$. It requires all three classifiers (CT, pathology, joint) to make correct predictions. During testing, the output from the joint classifier is treated as the final prediction.

\section{Experiments}

\subsection{Dataset and evaluation metrics}
\subsubsection{Dataset}
The dataset used in our experiments consists of CT volumes and H\&E stained histopathology whole slide images of 30 subjects, in particular one CT volume and one slide image per subject. All data are private data from our collaborative partner and are de-identified. The ground-truth fibrosis stage and NAS scores are provided by a pathologist with manual examination on the WSIs. We randomly split the 30 patients into three groups and perform 3-fold cross validation in all experiments.

The fibrosis stage ranges from 0 to 4. The total NAS score is made up of three individual scores - NAS steatosis, NAS lobular inflammation and NAS ballooning, which have 4, 4 and 3 different values, respectively. The original score distribution in the 30 subjects is shown in the Table~\ref{tab:score}.

\subsubsection{Label generation}
The original stage/scores cannot be used for training because the number of patients in some stages/scores are too small (e.g., 3 patients in fibrosis stage 3, 2 patients in NAS steatosis score 0). Based on our collaborating clinical doctor's input, we divide the fibrosis stages into three classes, NAS steatosis scores into two classes, NAS lobular into three classes and NAS ballooning into three classes. The distribution of the new classes are shown in Table 1 (`Combined' row).
In each combined class, there are relatively enough data for training.

\subsubsection{Evaluation metrics}
We use the Area Under ROC Curve (AUC) to evaluate the classification performance in our experiments. The ROC curve is created by plotting the true positive rate (TPR) against the false positive rate (FPR) at various threshold settings. AUC tells how much a model is capable of distinguishing between two classes. We also compute the 95\% confidence interval for AUC using the bootstrapping method with 1000 iterations. For the two-class problem (NAS steatosis), AUC values are averaged over the three folds to give the mean AUC. For the three-class problems (NAS lobular, NAS ballooning and fibrosis),  the AUC value of each individual class is computed by treating the other two classes as one class. The AUC of each fold is the average of AUC values of the three classes. The final mean AUC of an experiment is the average of three folds.

\begin{table*}[t]
\centering
\caption {Mean AUC values of fibrosis stage and NAS scores prediction using different methods (Three-fold cross validation).}
\begin{tabular}{|c |l |l |l |l |}
\hline
Method           & Fibrosis & NAS steatosis & NAS lobular  & NAS ballooning  \\ \hline
CT & 76.35$\pm$15.77 &	67.88$\pm$8.85 &	61.82$\pm$14.19	& 64.38$\pm$16.29 \\
H\&E & 83.85$\pm$9.68 &	94.04$\pm$10.33 &	64.85$\pm$4.8	& 74.63$\pm$17.19 \\
Mid-fusion Single-loss & \textbf{89.84$\pm$7.92}	& \textbf{95.95$\pm$7.02} &	\textbf{74.46$\pm$6.4} &	\textbf{77.19$\pm$13.36}  \\
Mid-fusion Multi-loss & 85.74$\pm$11.83 &	92.10$\pm$10.35 &	64.89$\pm$4.12 &	72.11$\pm$12.43 \\
Late-fusion Single-loss & 86.29$\pm$4.23 & 	91.12$\pm$3.51	& 65.83$\pm$5.13 &	73.78$\pm$20.89 \\
Late-fusion Multi-loss & 83.74$\pm$8.91 & 	85.06$\pm$15.8 &	64.54$\pm$15.06	& 66.18$\pm$20.11 \\ 
Simulation of Heinemann et al\cite{heinemann2019deep} & 59.06$\pm$10.11 & 	66.33$\pm$9.45 & 62.33$\pm$3.5	& 55.26$\pm$5.46 \\ \hline
\end{tabular}
\label{tab:combined}
\end{table*}

\subsection{Implementation Details}

We implement our method using the PyTorch~\cite{paszke2019pytorch} library.
 During training, the ResNet-18 without fc layer (the feature extractor) is initialized with pretrained weights from
 ImageNet~\cite{deng2009imagenet}. To avoid overfitting, only the last residual block is updated. For all experiments, the models are trained with the Adam optimizer for 30 epochs. The
 learning rate, batch size and weight decay are 0.0001, 4 and 0.01, respectively. The best model of the 30 epochs is selected for test. We have used the cross entropy loss in our experiments.

\subsection{Results and discussion}
\begin{figure*}[t]
    \centering
    \begin{minipage}{0.45\textwidth}
    \centering\includegraphics[width=\textwidth]{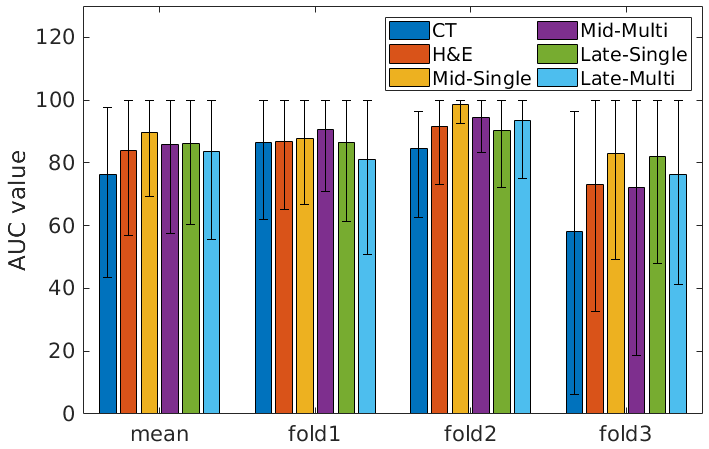} \\
    (a) Fibrosis
    \end{minipage}
    \begin{minipage}{0.45\textwidth}
    \centering\includegraphics[width=\textwidth]{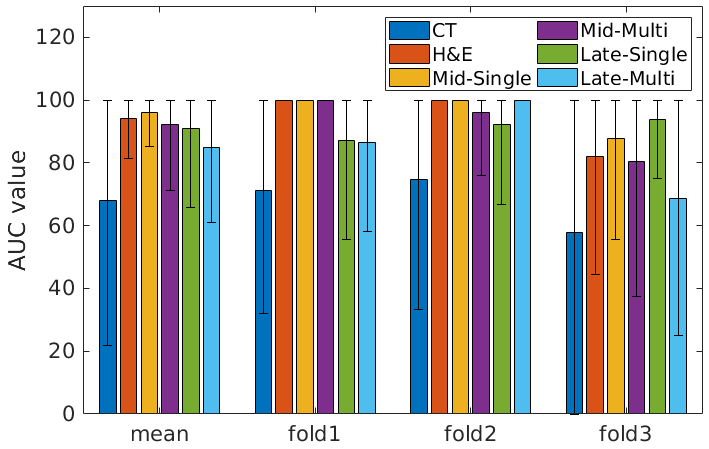} \\
    (b) NAS steatosis
    \end{minipage} \\
    \begin{minipage}{0.45\textwidth}
    \centering\includegraphics[width=\textwidth]{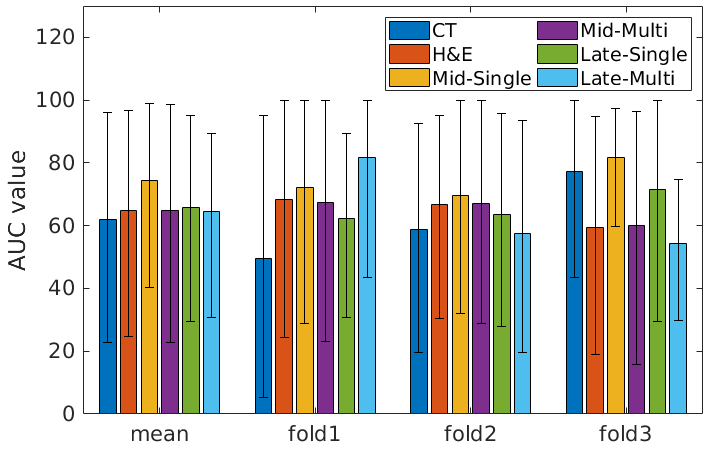} \\
    (c) NAS lobular
    \end{minipage}
    \begin{minipage}{0.45\textwidth}
    \centering\includegraphics[width=\textwidth]{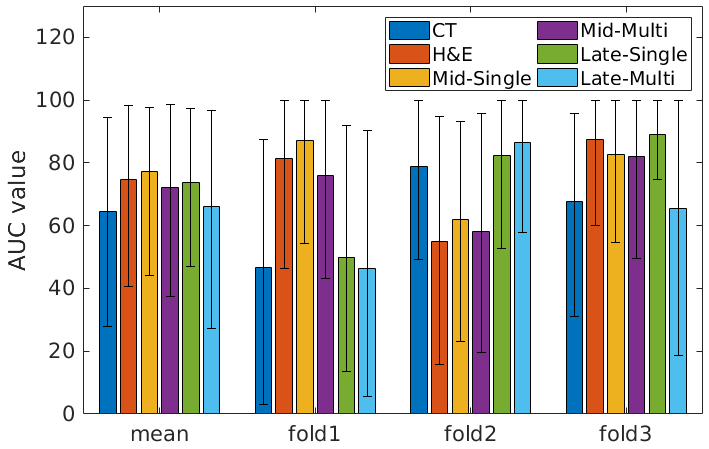} \\
    (d) NAS ballooning
    \end{minipage}
    \caption{AUC values and 95\% confidence intervals for different folds.}
    \label{fig:auc}
\end{figure*}

For each of the four prediction experiments (NAS steatosis, NAS lobular, NAS ballooning, fibrosis), we report the results using single type of data (CT or pathology), and both types of data on the four different architectures: mid-fusion with single loss, late-fusion with single loss, mid-fusion with multi-loss, late-fusion with multi-loss. The results are shown in Table~\ref{tab:combined} and Fig.~\ref{fig:auc}. We also compare our method with a latest related work~\cite{heinemann2019deep}. In that work, Inception-V3 (pretrained on ImageNet) backbone network is used to predict fibrosis and NASH score from rat/mouse liver pathology images. The results of their method on our data are shown in Table~\ref{tab:combined}.

\subsubsection{Comparison of results using single modality data}
When using only CT data, our baseline method can predict the fibrosis stage with 76.35 AUC score, but the results of NAS scores predictions are not good enough (AUC 61.82-67.88). The lobular imflammation and the ballooning are both microscopic structures, which are often analyzed in the pathology images. Therefore, using H\&E slides demonstrate a higher performance
score than using CT. Actually, H\&E outperforms CT on all four tasks, because H\&E images have more details about the cell and tissue structure.

\subsubsection{Comparison of results using multi-modality data}
The multi-modality based architecture (Mid-fusion with single-loss) achieves better results than any of the single modality (CT or H\&E) based architecture. The mean AUC values on fibrosis, NAS lobular, NAS ballooning are better than those in single modality. The NAS steatosis result is similar as that of H\&E because they are pretty high, and the room for improvement is limited. As for NAS lobular and NAS ballooning, the fused model can learn some useful features from CT data, although it is hard for diagnostic radiologists to find them in CT data. These results prove that the combination of CT and pathology data is indeed beneficial to train a more robust model.
\paragraph{Mid-fusion VS. late-fusion}
The results using mid-fusion strategy are better than those of late-fusion for both types of loss functions. Mid-fusion gathers the local features from each type of data while late-fusion combines the global features. The local features have more information about the images than global features, which could be the reason that mid-fusion works better than late-fusion.

\paragraph{Single-loss VS. multi-loss}
Whatever the feature fusion strategy is, single-loss achieves higher AUC value than multi-loss, indicating that the learning of single modality branches may have negative effects on the joint classifier. The overall performance may be improved if we use a more sophisticated method to adjust the weights of different branches. This is a topic of future work.

\subsubsection{Comparison to the latest related work}
As we can see from the table, Heinemann's\cite{heinemann2019deep} work performs worse on our dataset for both fibrosis and NASH scoring.
One of the reasons for the poor performance of Heinemann's work on our dataset could be due to the slightly different scoring procedure used in their work. Heinemann's work gives individual class labels to each patch from a single WSI slide which means two different patches from the same WSI slide can potentially get two completely different class labels. Secondly, the rule to determine the fibrosis and NASH score of a patch also varies slightly from the macroscopic scoring. An example is the scoring of NASH ballooning where scores are given based on the rule - if the patch does not have a ballooning cell then its class label is 0, else the class label is 1 which is a little different from the macroscopic scoring where classes are assigned based on the presence of (1) None (2) Few or (3) Many ballooning cells. In our work, we have patient level fibrosis and NASH scores. All the patches from a patient have the same fibrosis and NASH score as the patient's WSI slide.

\section{Conclusion and Future work}
This work explores the use of deep learning techniques to predict NAS scores directly from CT images, and to combine data from two different modalities for improving the estimation of fibrosis stage and the prediction of NAS score. Based on our positive results, in the future we plan to explore how this method scales when using more than two categories of data and also how it performs with  other types of data, e.g., ultra sonograpy, MRI, MRE. It will be better for clinical applications if we can increase performance using several types of non-invasive imaging data, such as CT and MRI. We also plan future research on how the weights of different losses will affect the final prediction result.

\bibliographystyle{unsrt} 
\bibliography{refs}

\vspace{12pt}

\end{document}